\documentclass[aps,twocolumn,floats,pre]{revtex4-1}
\usepackage{graphics,graphicx,epsfig}
\usepackage{amssymb,color,xcolor}
\usepackage{epsf,epstopdf,wrapfig}
\usepackage{mathtools}
\usepackage{placeins}

\newcommand{\beq}{\begin{equation}}
\newcommand{\eeq}{\end{equation}}
\newcommand{\beqn}{\begin{eqnarray}}
\newcommand{\eeqn}{\end{eqnarray}}
\newcommand {\e}[1]{\mathrm{~#1}}

\newcommand{\nn}{\nonumber}
\newcommand{\CO}{(Color online)\space}
\newcommand{\Cost}{\mathcal{M}}
\newcommand{\gmean}{\langle g \rangle}

\definecolor{darkgreen}{rgb}{0.0,0.33,0.0}
\definecolor{midgreen}{rgb}{0.0,0.5,0.0}

\begin{document}

\title{Extending the dynamic range of transcription factor action by translational regulation}
% Possible alternative title (TRS): ``Translational regulation extends/widens the dynamic range of transcription factor action''

\author{Thomas R. Sokolowski,\footnote{tsokolowski@ist.ac.at}$^1$ Aleksandra M. Walczak,$^2$ William Bialek,$^3$ and Ga\v{s}per Tka\v{c}ik\footnote{gtkacik@ist.ac.at}$^1$}

\affiliation{$^1$Institute of Science and Technology Austria, Am Campus 1, A-3400 Klosterneuburg, Austria\\
$^2$CNRS-Laboratoire de Physique Th\'eorique de l'\'Ecole Normale Sup\'erieure, 24 rue Lhomond, 75005 Paris, France\\
$^3$Joseph Henry Laboratories of Physics, Lewis--Sigler Institute for Integrative Genomics, Princeton University,
Princeton, New Jersey 08544, USA}

\date{\today}

\begin{abstract}
A crucial step in the regulation of gene expression is binding of transcription factor (TF) proteins to  regulatory sites along the DNA.  But transcription factors act at nanomolar concentrations, and noise due to random arrival of these molecules at their binding sites can severely limit the precision of regulation.  Recent work on the optimization of information flow through regulatory networks indicates that the lower end of the dynamic range of concentrations is simply inaccessible, overwhelmed by the impact of this noise.  Motivated by the behavior of homeodomain proteins, such as the maternal morphogen Bicoid in the fruit fly embryo, we suggest a scheme in which transcription factors also act as indirect translational regulators, binding to the mRNA of other transcription factors.  Intuitively, each mRNA molecule acts as an independent sensor of the TF concentration, and averaging over these multiple sensors reduces the noise.  We analyze information flow through this new scheme and identify conditions under which it outperforms direct transcriptional regulation.  Our results suggest that the dual role of homeodomain proteins is  not just a historical accident, but a solution to a crucial physics problem in the regulation of gene expression.
\end{abstract}

\maketitle

\section{Introduction}
Cells control the concentration of proteins in part by regulating transcription, the process by which mRNA molecules are synthesized from the DNA template.    Central to the regulation of transcription are the ``transcription factor'' (TF) proteins that bind to specific sites along the DNA and enhance or repress the expression of nearby genes.  Perhaps surprisingly, many TF molecules are present at very low concentration, and even at low total copy number \cite{Ptashne1986}. While it has been appreciated for many years that  low concentrations of biological signaling molecules must lead to significant noise levels \cite{Berg1977}, direct measurements of the fluctuations in gene expression have become possible only in the past fifteen years \cite{Elowitz2002}.   

Pathways for the regulation of gene expression can be seen as input-output devices, with information flowing from input control signals (TF concentrations) to output behaviors (number of synthesized protein molecules).  While ``information'' usually is used colloquially in describing biological systems, the mutual information between input and output provides a unique, quantitative measure of the performance of these systems \cite{Tkacik+Walczak2011_Review,Tkacik+Bialek2014_Review}.  In the context of embryonic development, for example, the information (in bits) carried by gene expression levels sets a limit on the complexity and reproducibility of the body plans that can be encoded by these genes \cite{Bialek2012}.

Decades of work on neural coding provide a model for the use of information theory in exploring signaling processes in biological systems \cite{Bialek2012,Spikes1999}.  To exploit this concept, as a first step it is necessary to estimate the various information theoretic quantities from data on real systems, and for genetic regulatory networks that has been achieved only very recently.  There are estimates of the mutual information between the concentration of a transcription factor and its  target gene expression \cite{Tkacik2008_PNAS,Hansen2015}, the information that expression levels of multiple genes carry about the position of cells in the developing fruit fly embryo \cite{Dubuis2013_PNAS,Tkacik+Dubuis2015}, and the information that gene expression levels provide about external signals in mammalian cells \cite{Cheong2011,Selimkhanov2014}.  As a second step, we need to understand theoretically how the various features of the systems---the architecture of signal transmission, the noise levels, the distribution of input signals---contribute to determining information transmission.  In qualitative terms, the noise levels set a limit to information flow given a fixed maximum signal level, and thus understanding information transmission is intimately connected to the question of how the cell can maximize the information conveyed by a limited number of molecules produced and transported stochastically \cite{Ziv2007,Tkacik2008_PRE,Tkacik2009_Info,Walczak2010,Tkacik2012,Sokolowski2015,Rieckh2014,deRonde2010,Tostevin2009,deRonde2011,deRonde2012}; completing the circle, this problem is directly analogous to the ``efficient coding'' problem in neural systems \cite{Barlow1961}.  As emphasized in Ref.~\cite{Tkacik+Bialek2014_Review}, information theoretic ideas can thus be used as tools for the quantitative characterization of biological systems, but there is also the more ambitious goal of building a theory in which the behavior of real neural, genetic, or biochemical networks could be derived, quantitatively, from the optimization of information flow.

Transmitting maximum information with a limited number of molecules requires regulatory networks to embody strategies for minimizing the effects of noise.  Importantly, there are (at least) two contributions to the noise \cite{Tkacik2008}, and optimal networks find a balance between these.  The more widely appreciated component of noise in transcriptional regulation comes from the stochastic birth and death of the synthesized protein and mRNA molecules \cite{Swain2002}, which we refer to as ``output noise.''  But there is also noise at the input of the regulatory process, from the random arrival of transcription factor molecules at their binding sites.  We can think of transcriptional regulation as mechanisms for sensing the concentration of TFs, connecting the analysis of ``input noise'' to the broader problem of limits on biochemical signaling and sensing \cite{Bialek2005,Bialek2008,Tkacik2009_Diff,Endres2008,Kaizu2014,Paijmans2014,tenWolde2015,Bicknell2015}, first studied in the context of bacterial chemotaxis \cite{Berg1977}.   Changing the shape of input-output relations, both through cooperativity and through feedback, changes the balance between input and output noise, thus rendering the optimization of information flow a well--posed problem even with very simple physical constraints on the total mean number of molecules  \cite{Tkacik2009_Info,Walczak2010,Tkacik2012}.

Central to any account of noise reduction is the effect of averaging.  In the context of transcriptional regulation, there is averaging over time as molecules accumulate, averaging over expression levels of multiple genes that are regulated by the same TF, and  
averaging over space as molecules diffuse between neighboring cells or nuclei, e.g., in a developing embryo \cite{Gregor2007b,Erdmann2009,Sokolowski2012,Sokolowski2015,Tikhonov2015} or organoid \cite{Mugler2015}.  In the (typical) case where one TF targets multiple genes, there is a regime where information transmission is optimized by complete redundancy in the response of these targets, and another regime in which the concentrations for activation or repression of the targets are staggered so as to ``tile'' the dynamic range of inputs \cite{Tkacik2009_Info}.  But, even as we consider networks with increasing numbers of targets, the optimal strategy is to insert the additional genes into the high concentration end of the input range, leaving the lower part of the dynamic range largely unused for information transmission.  In effect, the input noise at low concentrations is too high for reliable signaling, and the solution is simply to avoid this regime.  

Here we explore a scenario that allows transcription factors to recover access to the low end of their dynamic range.  Since these proteins bind to DNA, it is plausible that they could also bind to mRNA, thereby regulating translation; this is known to happen in the large class of homeodomain proteins \cite{Niessing2000,Rivera-Pomar1996,Dubnau1996}.  Intuitively, each mRNA molecule could act as an independent sensor of the TF concentration, and averaging over these multiple sensors could reduce the input noise and thereby allow for more effective information transmission at low TF concentrations.  To develop this intuition, we first consider a model of the ``direct transcriptional regulation'' (DTR) scheme, in which a TF concentration is read out by $M$ binding sites on the same promoter (Section \ref{secDTR}); we subsequently generalize the model to a more complicated ``indirect translational regulation'' (ITR) scheme, in which the redundant readout function is served by $M$ cytoplasmic mRNA molecules (Section \ref{secITR}). We compare the two regulation mechanisms by computing the maximum information flow in each as a function of the input noise magnitude and other determinants of the information flow (Section \ref{secComparison}). We conclude by discussing a biologically relevant example from early \textit{Drosophila} development (Section \ref{secDiscussion}).

\section{Averaging over neighboring regulatory regions in direct transcriptional regulation}
\label{secDTR}
\begin{figure}
\centering
\includegraphics[width = 3.5in]{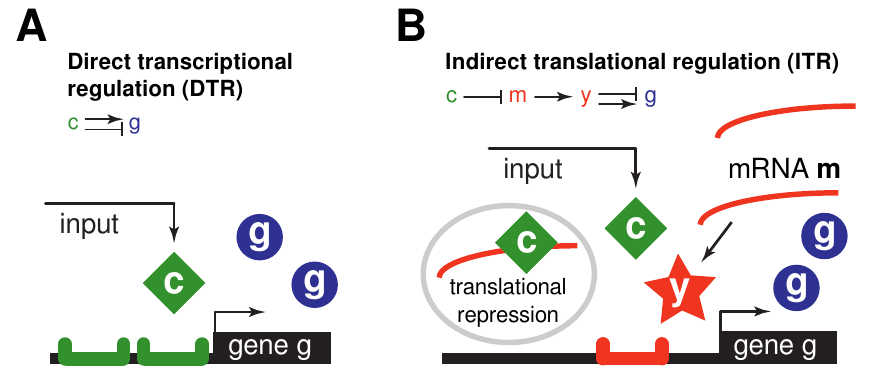}
\caption{\CO {\bf Schematic comparison of direct transcriptional regulation (DTR) and indirect tranlational regulation (ITR) schemes.} {\bf (A)} In direct transcriptional regulation (DTR), activator (or repressor) TFs, depicted as green squares and present at concentration $c$, interact with (potentially multiple, not necessarily identical) TF binding regions to activate (repress) expression of the regulated gene $g$. {\bf (B)} In the indirect translational regulation (ITR) scenario, input molecules (green squares) bind to mRNA $m$ of protein $y$ (red chain) to make the mRNA unaccessible for translation (gray oval). Translation can proceed from unbound mRNA molecules, giving rise to proteins $y$ (red stars). These proteins act as repressors (or activators) for gene $g$; the overall mapping from $c$ to $g$ is thus activating (repressing) in both scenarios.}
\label{f1}
\end{figure}

The intuition behind the arguments of this work is that a cell can extract more information from low concentrations of transcription factors by averaging over multiple binding regions.  We expect that this will be realized by having the multiple binding regions on different mRNA molecules.  As a motivating exercise, however, we can imagine that there are many regions for binding of the transcription factor at a single target near the gene being regulated, and that the expression of this gene depends on the average of the occupancies of these regions (see schematic in Fig.~\ref{f1}A); there are hints that such non--cooperative regulation by a cluster of binding regions may be realized in some cases \cite{Giorgetti2010}.  We expect that, with averaging over $M$ binding regions, we should find a $\sqrt{M}$ reduction in noise levels, and our goal here is to exhibit this explicitly, as well as to understand the conditions for this reduction to be achieved.  These results will provide a guide to the more complex case of ``indirect translational regulation'' (ITR), introduced in Sec.~\ref{secITR}.  The calculational framework we use here is based on our previous work \cite{Tkacik2009_Info,Walczak2010,Tkacik2012,Tkacik+Walczak2011_Review}.

We write the expression level of the single target gene as $g$, and if expression is controlled by the average of multiple nearby regulatory regions then the dynamics are of the form
\begin{equation}
{{dg}\over{dt}} = r \left[ {1\over M}\sum_{i=1}^M f_i (c)\right] - {1\over\tau} g + \xi ,
\label{modelA}
\end{equation}
where $r$ is the maximal rate of synthesis, $1/\tau$ is the rate at which the gene products are degraded, and $\xi$ is a Langevin noise source (zero-mean white noise).  In this model there is a single transcription factor, at concentration $c$, that controls expression.  We assume $\tau$ to be the longest time scale in the problem, thus setting the averaging time for all noise sources in the regulatory pathway. As described more fully in Refs.~\cite{Tkacik2008_PNAS,Tkacik2008_PRE,Tkacik2009_Info,Walczak2010,Tkacik2012,Sokolowski2015}, we can think of the regulatory mechanism as propagating information from $c$ to $g$, and this information transmission is a measure of the control power achieved by the system.  

In the simplest case, each region harbors just one binding site, and the contribution of that site to the activation of gene expression is determined by its equilibrium occupancy ${\bar n}_i\in[0,1]$; then we have
\begin{equation}
f_i(c) = {\bar n}_i (c) = {c\over{c+K_i}},
\end{equation}
where $K_i$ is the binding constant or affinity of site $i$ for the transcription factor. Alternatively, each region, corresponding to a regulatory sequence along the DNA, could be a tight cluster of binding sites that act cooperatively, so that
\begin{equation}
f_i(c) = {{c^{H_i}}\over{c^{H_i} + K_i^{H_i}}},
\label{HillFunctionDTR}
\end{equation}
where $H_i$ is the Hill coefficient describing the cooperativity.  Note that in this parameterization, we can describe activators and repressors by the same equation, using positive and negative $H_i$, respectively.

The noise term $\xi$ should include many different microscopic effects.  There is noise in the synthesis and degradation of the gene product (output noise), and there is noise in the arrival of the transcription factor at its target site (input noise).   As in Refs.~\cite{Tkacik2008_PRE,Tkacik2009_Info,Walczak2010,Tkacik2012,Sokolowski2015}, we describe the output noise as a birth--death process, and subsume several complexities by assuming that we are counting ``independent events'' without making a detailed commitment about their nature (e.g., whether the mRNA or protein molecules are independent, or if the truly independent events are bursts of transcription \cite{Golding2005}); we will, however, consider these aspects in more detail in the ITR model (Sec.~\ref{secITR}).

For the input noise, there is a minimum level set by the Berg--Purcell limit \cite{Berg1977,Bialek2005,Kaizu2014,Bicknell2015}, which is equivalent to a variance in the concentration of the transcription factor
\begin{equation}
\sigma_{c, {\rm (in)}}^2 = {c\over{D_c \ell_c \tau}} \Phi_B ({\vec n}(c)) , \label{innoise}
\end{equation}
where $D_c$ is the diffusion constant of the transcription factor, $\ell_c$ is the effective linear dimension of the binding region, $\tau$ is the integration time over which noise is averaged, and the ``occupancy factor'' $\Phi_B ({\vec n}(c))$ is a function of the average occupancy that depends on molecular details, with $B$ denoting the number of binding sites per regulatory region. In particular, for a single binding site with equilibrium occupancy $\bar{n}_1$, we expect $\Phi_1=(1-\bar{n}_1)^{-1}$ \cite{Kaizu2014,Paijmans2014}; for a cluster of sites in the limit as their number grows large, such that a single site is never fully saturated, $\Phi_B\rightarrow \Phi_\infty \equiv 1$ \cite{Bialek2005,Berezhkovskii2013}.

We can cast both input and output noise into the Langevin form (cf. Eq.~\ref{modelA}), but we know from Refs.~\cite{Tkacik2008_PRE,Tkacik2009_Info,Walczak2010,Tkacik2012,Sokolowski2015} that, so long as they are not too large, these noise sources provide additive contributions to the variance of $g$.  We can find the effect of the input noise by ``propagating errors'' through the mean input-output relations, and then add to the output noise:
\begin{equation}
\sigma_g^2 = \bar g + \sum_{i=1}^M \left( {{\partial \bar g}\over{\partial f_i}}\right)^2 \left( {{\partial f_i}\over{\partial c}}\right)^2 \sigma_{c, {\rm (in)}}^2 .
\end{equation}
The first term is the Poisson output noise, with the variance equal to the stationary mean, which can be computed from Eq.~(\ref{modelA}):
\begin{equation}
{\bar g}(c) = {{r\tau}\over M}\sum_{i=1}^M f_i (c) .
\end{equation}
Note that since $f_i \in [0,1]$, the maximum mean number of output molecules is $N_{\rm max} = r\tau$.

%
%
%
%
%
%%%%%%%%%%%%%%%%
%%% FIGURE 1 %%%
%%%%%%%%%%%%%%%%

Let us now assume, for simplicity, that all regulation functions $f_i$ are identical:  all $H_i = H$, all $K_i = K$, and hence all  $f_i(c) \equiv f(c) = {\bar n}(c)$. The total noise in gene expression in a model with $M$ identical binding regions then reads:
\begin{equation}
\sigma_g^2= \bar{g} + M\left(\frac{r\tau}{M}\right)^2\left(\frac{\partial f}{\partial c}\right)^2 \frac{c}{D_c \ell_c \tau} \Phi_B . \label{mnoise1}
\end{equation}
Introducing, consistently with our previous work, a dimensionless concentration unit $c_0 = N_{\rm max}/D_c \ell_c \tau$, and measuring expression levels $g$ in units of maximal induction $N_{\rm max}=r\tau$, we observe that the mean expression is simply $\bar{g}=f(c)$, and the noise can be written as
\begin{equation}
\sigma_g^2=\frac{1}{N_{\rm max}}\left[\bar{g} + \frac{\Phi_B}{M}c \left(\frac{\partial \bar{g}}{\partial c}\right)^2\right]. \label{modelAnoise1a}
\end{equation}
If the input concentration $c$ has a limited dynamic range, i.e., $c\in[0,C]$, where $C=c_{\rm max}/c_0$ is the maximal allowed concentration of the input in units of $c_0$, the relative importance of the two noise terms is set by $C$. For $C\gg 1$, it is possible to regulate the gene such that the input noise contribution [second term of Eq.~(\ref{modelAnoise1a})] is negligible compared to the output noise [first term of Eq.~(\ref{modelAnoise1a})]. For $C\ll 1$, the input noise is dominant and the output noise is negligible, unless $M$ is large. The balancing of these noise terms has been explored in detail in our previous work \cite{Tkacik2008_PRE,Tkacik2009_Info,Walczak2010,Tkacik2012,Sokolowski2015}.

Alternatively, the total noise at the output from Eq.~(\ref{modelAnoise1a}) can be mapped to an equivalent noise at the input through the slope of the input-output relation $\bar{g}'(c)$,
\begin{equation}
\sigma_c^2 = \frac{1}{N_{\rm max}}\left[\bar{g}\left(\frac{\partial \bar{g}}{\partial c}\right)^{-2} + \frac{{\Phi_B}}{M}c \right]. \label{modelAnoise1b}
\end{equation}

Equations~(\ref{modelAnoise1a},\ref{modelAnoise1b}) contain two differences compared to the single input/single output case reported in Ref.~\cite{Tkacik2009_Info}. First, the ``occupancy factor'' ${\Phi_B}$ is introduced by a refinement of the expression for the input noise \cite{Kaizu2014,Paijmans2014}; this change will not qualitatively influence our conclusions. Moreover, as mentioned above, the significance of this correction decreases as the number of binding sites per region increases \cite{Kaizu2014}. Second, the factor $1/M$ multiplying the input noise contribution suggests that the input noise can be decreased by averaging over multiple binding regions for the transcription factor.

If the transcriptional regulatory apparatus really were driven to strongly suppress the input noise, e.g., by a factor of order $10$, this would necessitate large $M$, and it is hard to imagine how $10^2-10^3$ binding regions could be packed into a linear regulatory section on the DNA. One difficulty with this is that there is no plausible molecular machinery which could read out the average occupancy of so many regions. The other, more fundamental difficulty is that, due to close packing on the DNA, such regulatory regions would interact and thus fail to provide independent concentration measurements, likely negating the apparent benefits of input noise averaging. This effect is well known since the original work of  Berg and Purcell in the context of chemoreception \cite{Berg1977}.

In the following section we will show that translational regulation implements an input noise reduction mechanism that is conceptually identical to that of $M$ binding regions, while automatically removing the two associated problems discussed above. We will compare the noise reduction in the ``indirect translational regulation'' (ITR) mechanism to the ``direct transcriptional regulation'' (DTR) case, which we define as the simple scheme using $M=1$ regulatory element with noise given by Eqs.~(\ref{modelAnoise1a},\ref{modelAnoise1b}).

\section{Indirect translational regulation}
\label{secITR}
In the indirect translational regulation (ITR) scenario, the gene $g$ is not regulated directly, but through an intermediate step. Let us assume that TF $c$ translationally represses the mRNA $m$ of a protein whose copy number we will denote by $y$; this protein acts as a repressing (or activating) TF for the output protein $g$, as depicted in Fig.~\ref{f1}B. As a result, the end transformation of inputs $c$ to outputs $g$ is again activating (or repressing), and can be compared to the respective direct transcriptional regulation pathway (see Fig.~\ref{f1}).

\pagebreak
One possible reaction scheme for indirect translational regulation consists of the following system of equations:
\begin{eqnarray}
\frac{dm}{dt}  &=& r_m - \frac{1}{\tau_m}m - k_+c m+ k_- b \label{system1}\\
\frac{db}{dt} & = &k_+ c\; m - k_-b - \frac{1}{\tau_m}b \\
\frac{dy}{dt}&=& r_y m - \frac{1}{\tau_y}y \\
\frac{dg}{dt}&=& r f(y/\Omega) - \frac{1}{\tau}g.\label{system2}
\end{eqnarray}
Here, the mRNA of the intermediary gene $y$ is produced at rate $r_m$ and degraded at rate $\tau_m^{-1}$. It can be bound by the input transcription factor at rate $k_+ c$, and unbound at rate $k_-$. The variable $m$ tracks the number of unbound mRNA from which translation can proceed; $b$ tracks the repressed (bound) mRNA number. Translation of the unbound mRNA occurs at rate $r_y$ and the protein is degraded with rate $\tau_y^{-1}$. Finally, $y$ controls the expression of $g$, as the input $c$ does in the DTR scenario, through a regulatory function $f$: $g$ proteins are expressed with maximal rate $r$, and degraded with rate $1/\tau$, by assumption the slowest time scale in the problem. Here $m$, $b$ and $y$ are given as absolute molecule counts, $c$ is given in concentration units. The function $f(\cdot)$ is defined to take concentration as input to parallel the DTR scenario; the expression level of the intermediary gene $y$ therefore must be divided by the relevant reaction volume, $\Omega$, when inserted into $f$. Equations~(\ref{system1}-\ref{system2}) still have to be supplemented by the associated Langevin noise terms; we analyze the noise in detail below.

By solving Eqs.~(\ref{system1}-\ref{system2}) in the steady state, we obtain the average levels of signaling molecules:
\begin{eqnarray}
\bar{m} &=& \frac{r_m\tau_m}{1+\frac{c}{K_c}} \nonumber \\
\bar{y} &=& r_y\tau_y \bar{m} \nonumber \\
\bar{g} &= &r\tau f\left(\frac{\bar{y}}{\Omega}\right). \label{ss}
\end{eqnarray}
Here, $K_c = \frac{1+k_-\tau_m}{k_+\tau_m}$, and in the limit of fast binding and unbinding ($k_-\tau_m\gg 1$) this reduces to $K_c\approx k_-/k_+$,  which is akin to the familiar form for the dissociation constant. As before, we define $N_{\rm max} =r\tau$ to be the maximum number of molecules at the output. Analogously, let $M=r_m\tau_m$ be the steady state number of mRNA, either active (unbound by $c$) or repressed (bound by $c$), i.e., $M=\bar{m}+\bar{b}$.

To compute the noise in gene expression at steady state, we consider the following noise sources: (i) shot noise due to the production and degradation of $g$; (ii)  shot noise due to the production and degradation of the protein $y$; (iii)  shot noise due to production and degradation of the mRNA $m$; (iv) diffusion noise due to random arrival of $y$ molecules at the promoter of $g$; and (v), diffusion noise due to random arrival of $c$ molecules at the mRNA of $y$. Only two of these noise sources [(i) and (v)] arise from directly analogous processes in the DTR scheme. The additional sources reflect the increased complexity of the ITR scheme, suggesting that ITR will only be beneficial over DTR when the reduction in the input noise of $c$ (v) is large enough to compensate for the effect of the new noise sources. 

As in previous work (Refs \cite{Tkacik2008_PRE,Tkacik2009_Info,Walczak2010,Tkacik2012,Sokolowski2015}, but see also~\cite{Rieckh2014}), this analysis neglects switching noise, i.e., the noise due to stochastic transitions between different occupancy states of the promoter \cite{Bialek2005,Sanchez2011,Hornos2005,Walczak2005_BPJ,Walczak2005_PNAS,Mugler2009}. One reason for this is that such noise contributions depend on the exact molecular mechanisms operational at the promoter, which we do not know in detail. The other reason is that the noise contributions analyzed here represent the physical bounds due to finite number of molecules.  At sufficiently low concentrations, which are  the focus of interest here, these noise contributions will overwhelm the switching noise; more generally we can imagine that cells have evolved mechanisms that minimize these adjustable sources of noise, leaving the physically inevitable noise sources to dominate.
%\TRS{Not least, under noise reduction schemes based on spatial averaging super-Poissonian switching noise behaves qualitatively similarly to input noise \cite{Erdmann2009,Sokolowski2012}, and we can expect the same for the mechanism studied here.}
We will now analyze the effect of the different noise contributions on the steady state variance of the output $g$ in detail. 

{\it Birth-death noise sources (i)--(iii).} To correctly compute the birth-death (shot) noise contributions to the total noise at the output, it is instructive to consider the propagation of arbitrary shot noise sources in a generic signaling cascade of the form:
\begin{eqnarray}
\frac{dx_1}{dt}&=&r_1 - \frac{1}{\tau_1}x_1 + \xi_1 		\nonumber \\
\frac{dx_2}{dt}&=&r_2 f_2(x_{1}) - \frac{1}{\tau_2}x_2  + \xi_2 \nonumber\\
&\vdots&\nonumber\\
\frac{dx_n}{dt}&=&r_n f_n(x_{n-1}) - \frac{1}{\tau_n}x_n  + \xi_n ~, 
\label{gnoise}
\end{eqnarray}
where the shot noise spectra are 
\begin{equation}
\langle \xi_j(t) \xi_k(t')\rangle = 2\tau_j^{-1}\bar{x}_j \delta_{jk}\delta(t - t')
\end{equation}
in the steady state, with ${\bar x}_j = \langle x_j \rangle$ the stationary mean.
%The linearized Fourier fluctuation in the last component can be written as a sum, $\delta x_n = \sum_{i=1}^n \delta_i$, where (prime denotes derivative)
%
%\begin{equation}
%\delta_i = \frac{\xi_i}{(-i\omega + 1/\tau_i)}\prod_{q=i+1}^n \frac{r_q f_q'}{-i\omega + 1/\tau_q}.
%\end{equation}
%
Assuming that $\tau_n \gg \tau_j$ for all $j\neq n$, it can be shown by linearizing and Fourier transforming Eqs.~(\ref{gnoise}) (see Appendix~\ref{AppendixNoisePropForm}) that the total variance in $x_n$, to a good approximation, is $\sigma_{n}^2 = \sum_{j=1}^{n-1} \sigma^2_{n\leftarrow j} + \bar{x}_n$, where
\begin{equation}
\sigma^2_{n\leftarrow j} = \left[ \prod_{q=j+1}^n r_q\tau_q f_q'\right]^2 \times \left(\frac{\tau_j}{\tau_n}\right) \times \bar{x}_j ~. \label{gnoise1}
\end{equation}
Equation~(\ref{gnoise1}) is intuitive to interpret: shot noise entering the cascade at step $j$ has variance equal to the mean, $\bar{x}_j$, which gets filtered by the temporal averaging, $\tau_j/\tau_n$, over what is ultimately the slowest timescale in the problem, $\tau_n=\tau$, and is finally propagated through all subsequent stages of the signaling pathway, given by the gain factors and slopes of the input-output relations \cite{Paulsson2004}.

{\it Diffusion noise sources (iv) and (v).} Here we first note that the contribution of diffusion noise sources to the variance in the output can be generically written as
\begin{equation}
\sigma_{g\leftarrow x}^2 = \frac{x}{D_x\ell_x\tau}\Phi_x\left(\frac{d\bar{g}}{dx}\right)^2,
\end{equation}
where $x$ is the concentration of the diffusing species, $\Phi_x$ is a function of the internal state of the regulatory region to which $x$ is binding, $\ell_x$ is its linear extent, and the derivative acts on the steady-state transformation between $x$ and the mean output $\bar{g}$, given by Eqs.~(\ref{ss}).

For noise source (iv), the diffusive species is $y$, and the target is the transcription factor binding site controlling the expression of $g$. The relevant input-output relation through which the noise is propagated is $\frac{\partial \bar g}{\partial (y/\Omega)}=r\tau f'(y/\Omega)$; to allow for flexible regulation that can implement different input-output curves $f(y/\Omega)$, we will later assume that the expression is effected by many binding sites, resulting in $\Phi_y\simeq 1$.  

For noise source (v), the diffusive species is $c$, and the targets are mRNA $m$. Since each mRNA molecule acts as an autonomous ``detector'' for the concentration $c$, the relevant ``receptor occupancy'' is the probability of a single mRNA to be bound and thus repressed by $c$; denoting the average single-mRNA occupancy as $\bar{m}_1 \equiv \bar m / M$, we can write the diffusion noise for each single mRNA as:
\begin{align}
 \sigma^2_{m_1} &= \left( \frac{\partial \bar{m}_1}{\partial c}\right)^2 \frac{c~\Phi_c}{D_{cm} l_c \tau} 
		 = \left( \frac{1}{M}\frac{\partial \bar m}{\partial c}\right)^2 \frac{c~\Phi_c}{D_{cm} l_c \tau}  ,\label{minnoise}
\end{align}
where  $D_{cm}=D_c + D_m$, accounts for the fact that in the ITR scenario also the detectors can be mobile;  usually we can expect $D_c \gg D_m$, such that $D_{cm} \simeq D_c$, and in the following we  thus set $D_{cm} = D_c$. As in the direct regulation model, we can  add up the diffusion noise for the identical but independent $c$-detectors to obtain the diffusion noise in the total mRNA population: $\sigma^2_m = M \sigma^2_{m_1}$. Propagating this noise through the downstream input-output relations finally yields the expression for noise source (v):
\begin{align}
 \sigma_{g\leftarrow c}^2 &= \left( \frac{\partial \bar g}{\partial \bar m}\right)^2 M \left( \frac{1}{M}\frac{\partial \bar m}{\partial c}\right)^2 \frac{c~\Phi_c}{D_c l_c \tau}
\end{align}
Assuming a single TF binding site per mRNA, we can expect $\Phi_c=(1-\bar{b}/M)^{-1}$ \cite{Kaizu2014}, which we use in our subsequent calculations.

{\it Assembling all noise sources together.} Applying the above considerations to the ITR regulation sheme defined by Eqs.~(\ref{system1}-\ref{system2}), we can write down the steady state variance in the output as:
\begin{widetext}
\begin{equation}
\sigma_g^2 = \underbrace{\left.\bar{g}\vphantom{\bigg)^2}\right.}_{\rm (i)} + \underbrace{\left(\frac{r f'}{\Omega}\right)^2 \tau\tau_y \bar{y}}_{\rm (ii)} + \underbrace{\left(\frac{rf'\tau}{\Omega}\right)^2(r_y\tau_y)^2\left(\frac{\tau_m}{\tau}\right)\bar{m}}_{\rm (iii)} + \underbrace{\frac{\bar{y}}{D_y \ell_y \tau\Omega }\Phi_yr^2\tau^2f'^2}_{\rm (iv)} + \underbrace{\frac{c}{D \ell_c \tau} M \Phi_c\left(\frac{1}{M}\frac{\partial\bar{m}}{\partial c}\right)^2\left(\frac{\partial\bar{g}}{\partial\bar{m}}\right)^2}_{\rm (v)}. \label{e1}
\end{equation}
%
% As with Eq.~(\ref{mnoise1}) in the case of $M$ binding sites, the diffusive noise term for molecules of $c$ binding to mRNA (v), each mRNA separately counts as a concentration ``detector,'' so the corresponding input-output relation is that of the single mRNA occupancy $\bar{m}/M$ (hence division by $M$ in the parenthesis); since there is $M$ such detectors, the variance is multiplied by $M$. %%% version 2
%
%
\end{widetext}

Let us now choose a set of units that is natural for the ITR scheme and consistent with the DTR scenario. As before, we measure the output $\bar{g}$ in units of $N_{\rm max}=r\tau$ such that it falls into $[0,1]$, and we measure the concentration $c$  in units of $c_0$,
\begin{equation}
c_0 = \frac{N_{\rm max} }{ D_c\ell_c\tau}.
\end{equation}
In direct analogy, we choose a concentration unit for proteins $y$:
\begin{equation}
y_0=\frac{N_{\rm max}}{D_y \ell_y \tau}.
\end{equation}
Since the binding sites and diffusion constants for TFs $y$ and $c$ are in principle different, these units could be different, but we will later assume them to be similar. Let us also define:
\begin{equation}
y_{\rm max} = \frac{r_y\tau_y}{\Omega y_0},
\end{equation}
as the maximum (dimensionless) concentration of proteins $y$ expressed from a \textit{single} mRNA.

Binding and unbinding of $c$ to mRNA defines the first nonlinearity of the problem; denoting the average occupancy of a \textit{single} mRNA molecule by $h(c)$, we have
\begin{equation}
h(c) = \frac{K_c}{K_c + c}, \label{chill}
\end{equation}
where $c$ and $K_c = (1+k_-\tau_m) / (k_+\tau_m c_0)$ are both dimensionless. Note that $h(c) = \bar{m}/M$. The second nonlinear function is $f(y)$, determining the output expression level of $g$ (as in the DTR scenario); its argument is a dimensionless concentration of $y$, measured in units of $y_0$. We will explore the space of functions with a Hill form
\begin{equation}
f(y) = \frac{K^H}{y^H + K^H}, \label{yhill}
\end{equation}
where $K$ is also measured in units of $y_0$, and the Hill coefficient $H$ roughly corresponds to the number of $y$ binding sites on the promoter of $g$. Here again the sign of $H$ determines whether $y$ is an activator or repressor to $g$. In this work, we will focus on the case $H<0$ (meaning that $y$ activates $g$, while itself being repressed by the upstream input $c$, resulting in {\it overall} negative regulation of $g$ by $c$), and compare to the DTR model in which $g$ is repressed by $c$ as well [$H<0$ in Eq.~(\ref{HillFunctionDTR})].
%In this work, we will focus on {\it overall} negative regulation of $g$ by the upstream input $c$, corresponding to $H<0$ (meaning that $y$ activates $g$, while itself being repressed by $c$), and compare to the DTR model in which $c$ represses $g$. %%% previous version

\begin{widetext}
We can rewrite the noise in Eq.~(\ref{e1}) as the effective noise at the input,
\begin{equation}
 \sigma_g^2 = \left(\frac{\partial f}{\partial (\bar y/y_0)}\right)^2 \left(\frac{\partial (\bar y/y_0)}{\partial \bar m}\right)^2 \left(\frac{\partial \bar m}{\partial c}\right)^2 \sigma^2_c
	    = y_{\rm max}^2 M^2 f'^2 h'^2 ~ \sigma^2_c.
\end{equation}
After rearranging terms and writing all copy numbers and concentrations in the new units defined above, we obtain
\newcommand{\vph}{\vphantom{\frac{1}{y_{\rm max}^2}}}
\begin{equation}
\sigma_c^2 = \frac{1}{N_{\rm max}} \left[
    \underbrace{ \frac{1}{M^2} \frac{f}{f'^2} \frac{1}{h'^2} \frac{1}{y_{\rm max}^2} }_{\rm (i)}
  + \underbrace{ \frac{F}{M} \frac{h}{h'^2} \vph }_{\rm (ii)+(iii)} 
  + \underbrace{ \frac{1}{M} \frac{h}{h'^2} \frac{\Phi_y}{y_{\rm max}} \vph }_{\rm (iv)} 
  + \underbrace{ \frac{\Phi_c}{M}c \vph }_{\rm (v)} 
\right], \label{e2}
\end{equation}
where 
\begin{equation}
F=\frac{r}{r_y}(1+r_y\tau_m), \label{eqf}
\end{equation}
and $f$ and its derivative are evaluated at $\bar y/y_0 = y_{\rm max} \bar m = y_{\rm max} M h(c)$. As before, the input concentrations can vary across the range $c\in [0, C]$, where $C=c_{\rm max}/c_0$ is the dimensionless maximal concentration of the input. If gene expression rates for the target and intermediary proteins were similar, $r\sim r_y$, then $F\approx 1+ r_y\tau_m$, which is reminiscent of the Fano factor due to the ``burst size'' $r_y\tau_m$, the number of proteins of $y$ expressed on average from one mRNA; we will therefore refer to $F$ by ``Fano factor'' in the following. In a modified variant of the ITR model where mRNA are not continuously created and degraded to maintain an average of $M$ copies, but are present at a fixed total number of $M$ copies, $F=r/r_y\approx 1$.

\clearpage %%% Leave this inside the widetext environment for correct working! Otherwise we get a hanging formula...
\end{widetext}

It is useful to remind ourselves of the corresponding result in the DTR case, which is Eq.~(\ref{modelAnoise1b}) with $M=1$,
\begin{equation}
\sigma_c^2 = \frac{1}{N_{\rm max}}\left[
		 \underset{ {\rm (i')} }{ \underbrace{\bar{g}\bigg(\frac{\partial \bar{g}}{\partial c}\bigg)^{-2}} } 
	       + \underset{ {\rm (v')} }{ \underbrace{{\Phi_B} c \vphantom{\bigg)^{-2}}} } 
	     \right]. 
\end{equation}
We see that term (i) of the ITR case is equal to term (i$'$) in the DTR case, although this identification requires us to propagate the inputs through more layers of response in the ITR, hence a more complex expression. Similarly, term (v) of the ITR case, representing diffusive noise for $c$-molecules binding to mRNA $m$, is directly analogous to the diffusion noise term (v$'$) in the DTR case, but ITR reduces the input noise variance by a factor of $M$. 
%At equal $M$---while it is unclear how this could be implemented in vivo---the DTR scheme obviously would perform better, due to the new noise contributions from the intermediary regulatory steps arising in the ITR model, terms (ii), (iii) and (iv), that we will together refer to as ``ITR noise''.
%The crucial difference is the suppression of the effective noise in $c$ by a factor of $M$, which mimics the results obtained in the DTR scenario with multiple regulatory regions ($M>1$). 
While Eq.~(\ref{e2}) appears complicated, we can nevertheless estimate the relative magnitudes of different noise sources and assess their relevance. 

%In gene regulation the gain from introducing non-trivial noise-control mechanisms usually consists of suppressing super-Poissonian parts of the input noise; \TRS{in the following we will thus compare the scaling of the input noise terms (iv) and (v) relative to the output noise (i), and to the (propagated) shot noise from making intermediary compounds, (ii)+(iii), which we term ``translation noise''.} Note that in all scaling arguments involving the regulatory functions $h(c)$ and $f(y)
%$ and their derivatives, we will assume that the activation/repression thresholds $K_c$ and $K_y$ have the same scaling as the function arguments.

Let us first compare the relative magnitude of the two input-type noise sources. 
The scale of the ITR noise component due to finite number of $y$ molecules (iv) relative to the input noise in $c$ (v) is 
\begin{equation}
\frac{\rm (v)}{\rm (iv)}= y_{\rm max}\frac{ h'^2}{h}\frac{\Phi_c}{\Phi_y}c\sim\frac{y_{\rm max}}{C}.
\end{equation}
We expect that  $\Phi$ and regulatory functions $h$ are of order unity, and that the natural  scale of the concentration $c$ is given by $C$, the maximal dimensionless input concentration; then  the scale of the derivative of $h$ is $h'\sim C^{-1}$, leading to the final result. The importance of this term thus depends on the comparison of the maximal concentration of the intermediary $y$ proteins and the input proteins $c$. Clearly, if the intermediary proteins of $y$ are present at very low copy numbers, their diffusion noise will become limiting, and the ITR scheme will be ineffectual.

The scale of input noise (v) relative to the components of the ITR noise due to birth-death processes, (ii)+(iii), is given by
\begin{equation}
\frac{\rm (v)}{\rm (ii) + (iii)}=\frac{ h'^2}{h}\Phi_c\frac{c}{F}\sim \frac{1}{FC}.
\end{equation}
As we argue above, we should expect values of $F\geq 1$. The importance of noise sources (ii)+(iii) thus depends on the scale of $F$ relative to $C$, and in the regime of low input concentration, $C\ll 1$, noise sources (ii)+(iii) will be negligible for models with small transcriptional bursting for $y$ proteins, i.e., when $F\sim 1$.
%Following the same argumentation, an identical scaling is found for the magnitude ratio $\frac{\rm (iv)}{\rm (ii)+(iii)}$ with $C$ replaced by $y_{\rm max}$.

Finally, we can assess the relative scale of the input noise (v) with respect to the output noise (i):
\begin{equation}
\frac{\rm (v)}{\rm (i)}=M\frac{h'^2\; f'^2}{f} y_{\rm max}^2\Phi_c c\sim {\frac{H^2}{MC}}.
\label{eqScalingInToOut}
\end{equation}
Here, the regulation functions $f$ and $h$ are again of order unity, but the derivative of $f$ has a scale of $H/(M y_{\rm max} h)$, $H$ being the Hill coefficient of $f$. As a result, the dependence of the noise term (i) on $y_{\rm max}$ exactly cancels out; similarly, the $M$-dependence of term (i) cancels out exactly. This is to be expected: by increasing $M$, one can average away the input noise, but the magnitude of the output noise can not be reduced. The same is true for Eq.~(\ref{modelAnoise1b}), where the output noise contribution is not divided by $M$. This term is thus important as it must become limiting when $M$ grows large, with the relevant scale being set by $H^2/MC\sim 1$. The quadratic scaling with $H$ shows that steep regulation curves strongly amplify the noise on the input side.
%As before, identical scaling laws hold for the other input-to-output noise magnitude ratio $\frac{\rm (iv)}{\rm (i)}$ and the ratio $\frac{\rm (ii)+(iii)}{\rm (i)}$, with $C$ replaced by $y_{\rm max}$ and $1/F$, respectively.}

%
%
%
%
%
%%%%%%%%%%%%%%%%
%%% FIGURE 2 %%%
%%%%%%%%%%%%%%%%
\begin{figure*}[ht]
\centering
\includegraphics[height = 3in]{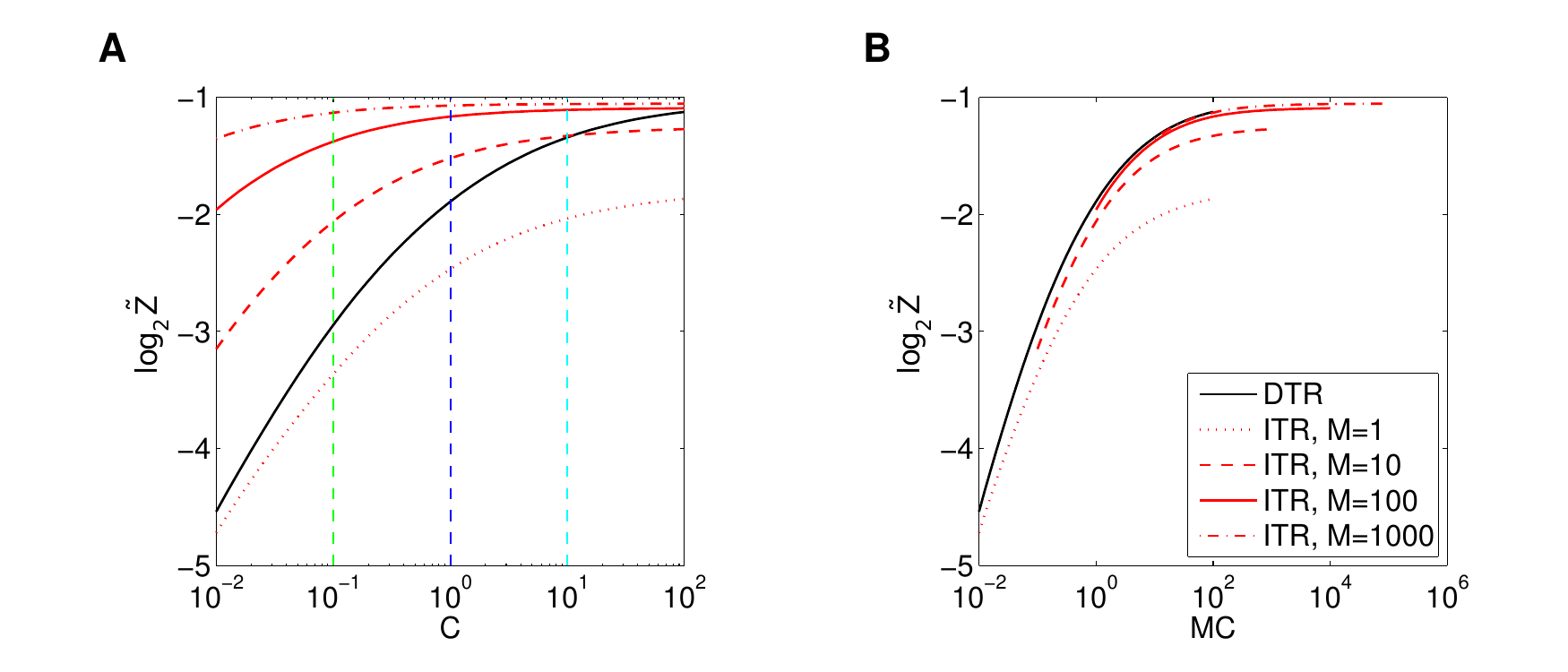}
\caption{\CO {\bf Optimal capacity of direct transcriptional regulation (DTR, black), compared to indirect translational regulation (ITR, red), as a function of the maximal input concentration $C$.}
{\bf (A)} Optimal capacity $\log_2\tilde{Z}$ vs. $C$ for various total amounts of translationally targeted mRNAs, $M$. Here and in subsequent panels we fix $N_{\rm max}$, the maximal output copy number, to a reference value $N_{\rm max}=1$ and show the resulting capacity; larger values of $N_{\rm max}$ simply shift all capacities upwards by an additive amount, as in Eq.~(\ref{c1}). The strength of ``ITR noise'' sources, (ii) + (iii) + (iv) of Eq.~(\ref{e2}), is set by (Fano factor) $F=1$ and (maximal concentration of the intermediary protein) $y_{\rm max}=10$. For $M>1$, ITR outperforms DTR at low $C$, but at high $C$ DTR can still reach higher capacities if $M$ is not sufficiently large (e.g., dashed red line for $M=10$). At $M=1$ (dotted red line), the ITR scheme cannot benefit from input noise averaging, yet the intermediary regulatory steps still contribute the ``ITR noise'' absent in DTR, causing DTR to be superior to ITR at all $C$.
{\bf (B)} Capacity curves for different values of $M$ in (A) collapse when plotted against the product of maximal input concentration and the number of mRNA targets, $MC$, as predicted by the scaling relation in Eq.~(\ref{eqScalingInToOut}). Increasing input noise by lowering $C$ thus can be compensated for by increasing $M$. The collapse is not perfect at high $MC$ and curves for different $M$ saturate at different capacity values because the strength of ``ITR noise'' is not negligible and their effect on capacity depends on $M$.
}
\label{f2}
\end{figure*}
%
%
%
%
%
%%%%%%%%%%%%%%%%
%%% FIGURE 3 %%%
%%%%%%%%%%%%%%%%
% (ex-figure 4)
\begin{figure*}[ht]
\centering
\includegraphics[height = 3in]{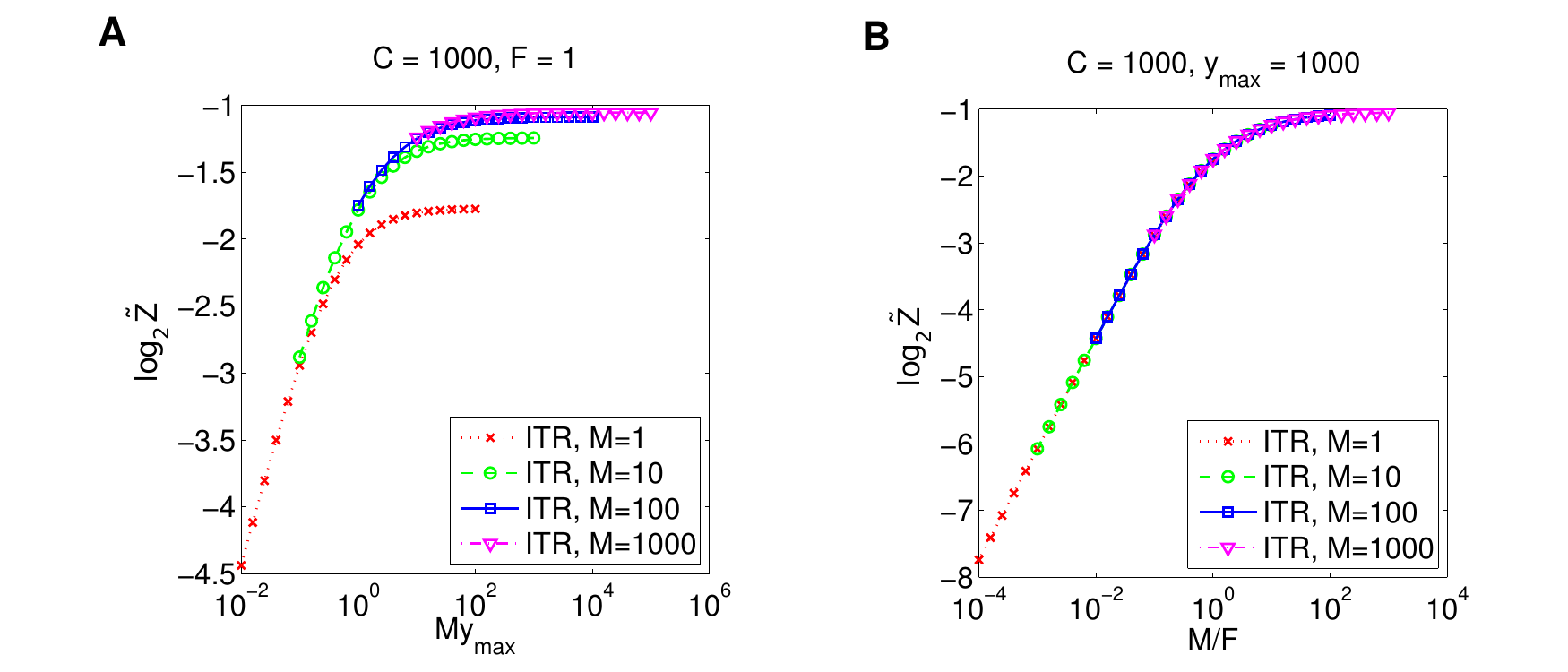}
\caption{
\CO {\bf Scaling of the optimal capacity with ``ITR noise'' parameters $y_{\rm max}$ and $1/F$.}
The optimal capacity $\log_2 \tilde{Z}$ for different fixed values of the mRNA number $M$ in the ITR model is plotted against ``ITR noise'' parameters properly rescaled by $M$, to observe the compensation between translational noise components and the input noise.
{\bf (A)} For the diffusion noise due to the intermediary protein $y$, the relevant parameter is $M \times y_{\rm max}$, where $y_{\rm max}$ is the maximal concentration of $y$.
{\bf (B)} For the shot noise due to the expression of intermediary mRNA and protein $y$, the relevant parameter is $M \times (1/F)$, where $1/F$ is the inverse Fano factor.
In both cases, we set the remaining noise sources to be as small as possible: $C=1000$ and $F=1$ for (A), $C=1000$ and $y_{\rm max}=1000$ for (B). A perfect collapse in (A), comparable to that in (B), could only be achieved for Fano factors $F\ll 1$, which are biologically unrealistic.}

\label{f3}
\end{figure*}
%
%
%
%
%
%%%%%%%%%%%%%%%%
%%% FIGURE 4 %%%
%%%%%%%%%%%%%%%%
% (ex-figure 3)
\begin{figure*}
\centering
\includegraphics[height=4.6in]{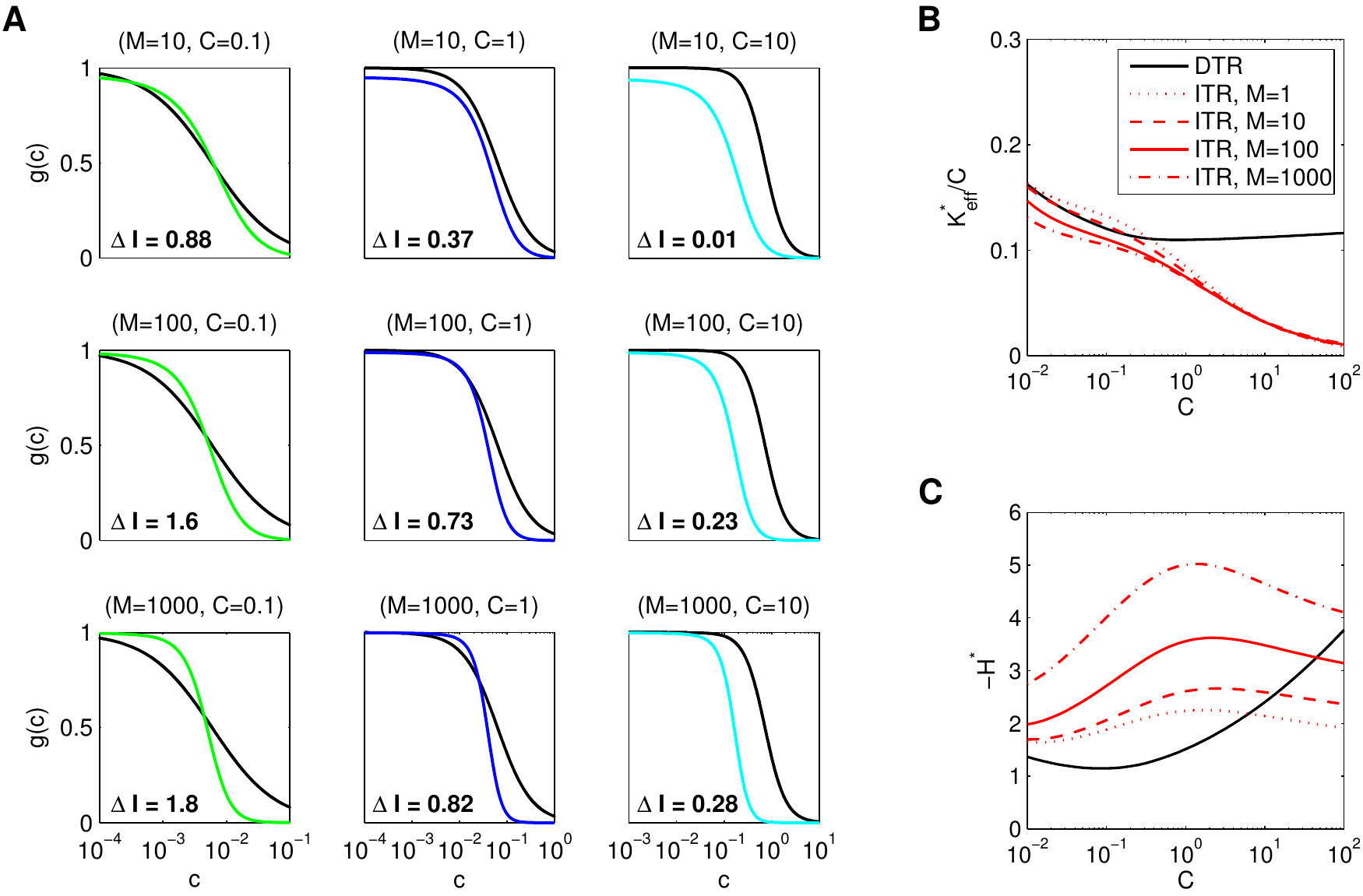}
\caption{
{\CO {\bf Optimal regulatory curves.} {\bf (A)} Optimal regulatory curves, $\bar{g}(c)$, for $C=0.1$ (left), $C=1$ (middle) and $C=10$ (right), and different values of $M$ in the indirect translational regulation (ITR) model (colored curves): $M=10$ (top), $M=100$ (middle), $M=1000$ (bottom). The direct transcriptional regulation (DTR) model is plotted in black for reference. Colors correspond to the vertical lines in Fig.~\ref{f2}. Here, $F=1$ and $y_{\rm max}=10$. $\Delta I$, the difference in capacity between ITR and DTR in bits, is shown in each panel.
{\bf (B, C)} Optimal regulation threshold, $K_{\rm eff}^*$, and the optimal Hill coefficient, $H^*$, that characterize the regulatory curves in (A), shown as a function of $C$. Since the overall regulation is repressive in either scheme, Hill coefficients are negative.
In the DTR model, $K_{\rm eff}\equiv K$; for ITR, $K_{\rm eff}=(K_c/K) M y_{\rm max} - K_c$.}}
\label{f4}
\end{figure*}

\section{Comparing optimal information flow in the two schemes}
\label{secComparison}
To compare the regulatory power of direct transcriptional regulation (DTR) and indirect translational regulation (ITR), we ask how much information can be transmitted through each scheme. If noise in gene expression is Gaussian, as we assumed here, then the response of the regulatory pathway is fully characterized at steady state by the distribution $P(g|c)=\mathcal{G}(g; \bar{g}(c), \sigma_g(c))$, where $\mathcal{G}$ is a normal distribution with the mean and standard deviation that can be computed from the Langevin equations Eqs.~(\ref{modelA},\ref{system1}-\ref{system2}). The mutual information, $I(c;g)$, between the regulator concentration, $c$, and the downstream expression level, $g$, is then given by:
\begin{equation}
I(g;c) = \int dc\; P_c(c) \int dg\; P(g|c)\log_2\frac{P(g|c)}{P_g(g)},
\end{equation}
where $P_g(g) = \int dc\; P_c(c) P(g|c)$. {Mutual information, a non-negative number measured in bits, tells us how precisely the input $c$ determines the output $g$, given that the noise places limits to the fidelity of this control. This quantity still depends on the distribution of input TF levels, $P_c(c)$, experienced by the regulatory pathway. It is possible to find the optimal distribution $P^*_c(c)$ that maximizes the information, and the maximum achievable information is then referred to as the (channel) capacity \cite{Tkacik2008_PNAS,Tkacik+Walczak2011_Review}. $P^*_c(c)$ is tuned to the noise properties of the regulation process, favoring the use of input concentrations at which the regulatory element responds with smaller noise over those inputs where noise is higher.

While finding the optimal input distribution and the corresponding maximal information $I$ is difficult in general, in the case of small noise, $\sigma_g\ll 1$, we previously derived a simple formula for the capacity \cite{Tkacik2008_PNAS}:
%
%\begin{eqnarray}
%I(c;g) &=& \log_2\frac{Z}{\sqrt{2 \pi e}} \label{c1} \\
%&\equiv& \log_2 \tilde{Z}+\frac{1}{2}\log_2(N_{\rm max}), \nonumber
%\end{eqnarray}
\begin{equation}
I(c;g) = \log_2\frac{Z}{\sqrt{2 \pi e}} \label{c1} 
\equiv \log_2 \tilde{Z}+\frac{1}{2}\log_2(N_{\rm max}),  
\end{equation}
where 
\begin{eqnarray}
Z &=& \int_0^C \frac{dc}{\sigma_c(c)}, \label{c2} \\
\tilde{Z} &=&Z(N_{\rm max}=1)/\sqrt{2\pi e}
\end{eqnarray}
and $\sigma_c$ is given either by Eq.~(\ref{modelAnoise1b}) for the case of direct transcriptional regulation, or by Eq.~(\ref{e2}) for the case of indirect translational regulation. $Z$ plays the role of the normalization constant in the distribution over inputs that maximizes $I(c;g)$; the optimal distribution is $P_c^*(c) = Z^{-1}\sigma_c^{-1}(c)$. The simple dependence of capacity on $N_{\rm max}$ in Eq.~(\ref{c1}) follows because $N_{\rm max}$ only enters the expression as a multiplicative prefactor to $\sigma_c$. 

The capacity given by Eqs.~(\ref{c1}, \ref{c2}) still depends on the regulatory parameters.  We view the maximum concentration of input molecules, $C$, not as a parameter but as a constraint; in the ITR case there is also a constraint on the number of mRNA molecules, $M$, and, less importantly, on the statistics of translation, captured by $F$. In the DTR scheme, the parameters that we can adjust in order to optimize information transmission are the dissociation constant ($K$) and the Hill coefficient ($H$) of the regulatory function $f$ in Eq.~(\ref{modelA}). For ITR, we have the same two parameters determining the properties of the regulatory function $f$ in Eq.~(\ref{system2}), plus an extra parameter, $K_c$, the dissociation constant for the repression function $h(c)$.   
%Given $C$ (and in the ITR case: $y_{\rm max}$, $M$, and $F$), we can maximize $I(c;g)$ over the regulatory parameters to obtain the channel capacity for every $C$, and the corresponding optimal regulatory curves, $\bar{g}(c)$. 
Since our focus is on the regime where input noise is limiting, $C\ll 1$, we fix $y_{\rm max}$ and $F$ such that ``ITR noise'' sources, relating to intermediary compounds, i.e., the mRNA and protein of $y$, are not dominant. The analysis of noise terms from the preceding section suggests that choosing $F=1$ and $y_{\rm max}=10$ will ensure that noise in the ITR scenario is dominated by the input noise in $c$ [(v)] and the output noise due to production and degradation of $g$ [(i)]. In this regime, the performance of the complex ITR regulatory scheme should approach the direct regulation using $M$ regulatory elements, Eq.~(\ref{modelAnoise1b}).

\subsection{Optimal solutions}
\label{secOptimalSolutions}

We mapped the optimal capacity as a function of $C$ for the DTR and ITR scenarios at various values for $M$, with the results shown in Fig.~\ref{f2}. For each point on the capacity curves, information was maximized with respect to the regulatory parameters ($\{K,H\}$ in case of DTR, $\{K,H,K_c\}$ in the case of ITR). As expected, increasing $M$ clearly enhances the capacity by additionally suppressing dominant input noise at $C<1$ relative to direct regulation. For $C>1$, the performance of ITR vs. DTR depends on the choice of $y_{\rm max}$ and $F$, which determine when additional noise sources specific to the ITR scheme  become important. 

At  very small $C\sim 10^{-2}$, the input noise clearly limits capacity in the DTR scenario. As we switch over to the ITR scenario with $M=10$, we expect the (dominant) input noise variance to drop by a factor of 10, leading to an increase in $I(c;g)$ of $\log_2\sqrt{10}\approx 1.7$ bits, roughly equal to the observed difference between the DTR curve and  the ITR curve for $M=10$. As $M$ is increased, this   scaling breaks down because the reduced input noise stops being the sole factor limiting information transmission, and the capacity curves flatten  as a function of $C$, saturating towards the limit where the transmission is limited only by output and ITR noise.  We draw attention to  the magnitude of these effects:  the capacity of real regulatory elements is in the range of one to several bits \cite{Tkacik2008_PNAS}, so that differences on the order of one bit are huge.

The value of $C$ where the capacity begins to saturate depends on $M$, because increasing $M$ is equivalent to increasing $C$ as predicted by the scaling relation of Eq.~(\ref{eqScalingInToOut}). To make this explicit we plotted the same capacity data against $MC$ instead of $C$, as shown in Fig.~\ref{f2}B. As expected, the curves collapse in the low-$MC$ regime, where input noise is dominant, and start to saturate at similar $MC$ values. However, they do not reach the same plateau at saturation, because the ITR noise contributions, (ii) + (iii) + (iv), are non-negligible.
% with the chosen values of the ``Fano factor'' $F$ and dynamic range of the intermediary protein, $y_{\rm max}$. 
To see that these contributions can be traded off against the decrease in input noise set by $M$ as well, we varied parameters $F$ and $y_{\rm max}$ at fixed $C=1000$ in Fig.~\ref{f3}, such that only one noise source remained non-negligible compared to the others. When the resulting capacity is plotted against the appropriately rescaled versions of the parameters, $My_{\rm max}$ and $M/F$, respectively, we again see a collapse of the capacity curves for different $M$, as predicted in Section~\ref{secITR}. % This time we set the two other control parameters ($C$ and $1/F$ in \ref{f4}A, $C$ and $y_{\rm max}$ in \ref{f4}B) to high values, to diminish the magnitude of the corresponding noise contributions. As in Fig.~\ref{f2}B and as predicted by the scaling law, the capacity curves 
%collapse for low values of the rescaled control parameters, and saturate when they reach high values.
In Fig.~\ref{f3}B the collapse is almost perfect, because $y$-input noise (iv) dominates whereas the respective other two noise sources are negligible, and the plateau thus is purely set by output noise. In Fig.~\ref{f3}A this is not possible because under realistic conditions the Fano factor $F\geq 1$, meaning that the ITR noise cannot be completely eliminated; however, we verified that a perfect collapse can also be achieved here by setting $F$ to unrealistically low values $F\ll 1$.

What do the optimal regulatory curves look like? Figure~\ref{f4}A shows that at low $C=0.1$, where capacity enhancement by translational regulation is largest, the lowered input noise in the ITR model allows for markedly steeper effective input-output curves $\bar g(c)$, especially for $M\gg 1$. At larger $C$, the difference in capacity between ITR and DTR decreases, as expected from Fig.~\ref{f2}, and the input-output curves converge towards similar effective Hill coefficients in both models, but ITR has a significantly  lower midpoint input concentration than DTR. These observations are confirmed by a more systematic analysis, presented in Figs.~\ref{f4}B and C, where we plot the optimal (effective) transition point $K_{\rm eff}^*$ and the optimal Hill coefficient $H^*$, respectively, as functions of $C$. 
% Note that--contrary to the general trend--at low $C$ and larger $M$ the transition point $K_{\rm eff}$ in the ITR model 
% is at slightly higher concentrations as in the DTR model;
%

%
%
%
%
%
%%%%%%%%%%%%%%%%
%%% FIGURE 5 %%%
%%%%%%%%%%%%%%%%
\begin{figure*}[ht]
\includegraphics[height = 3in]{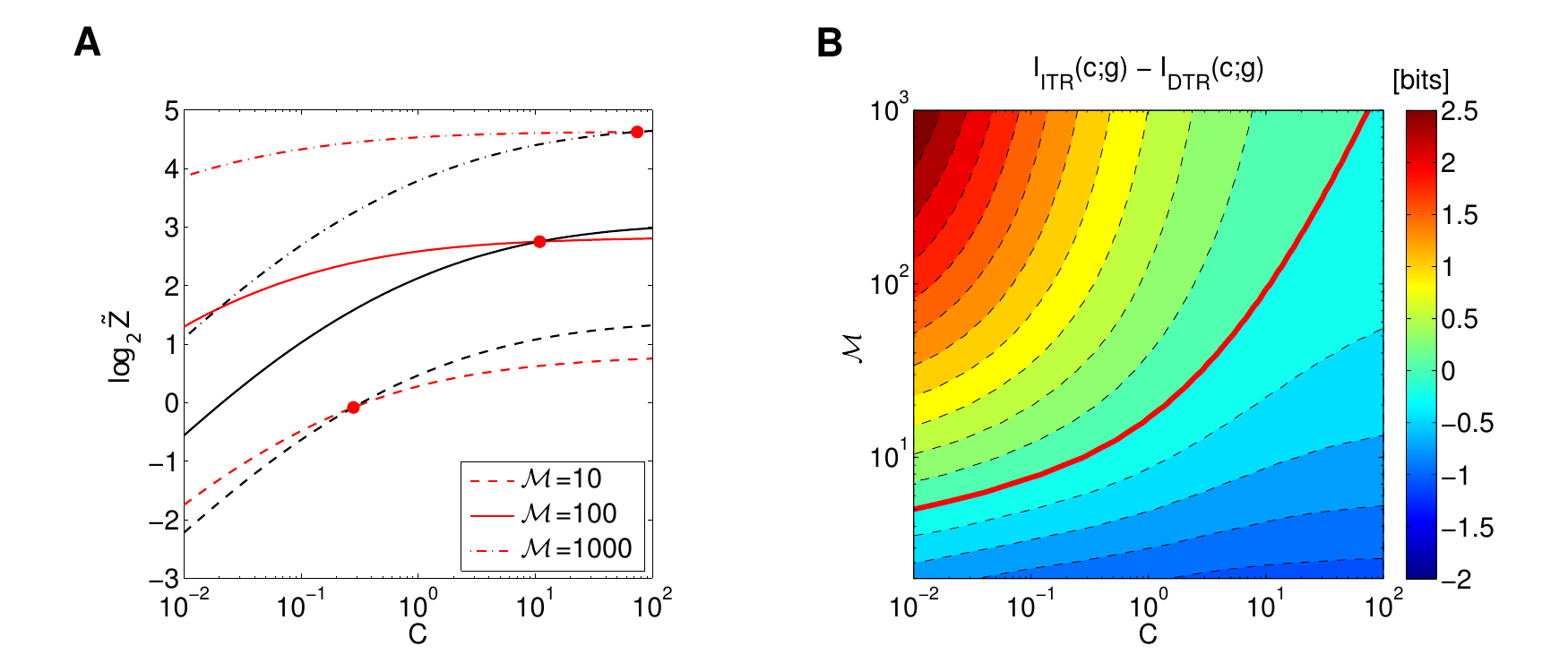}
\caption{\CO {\bf Direct transcriptional regulation (DTR) vs. indirect translational regulation (ITR) at equal resources.} 
{\bf (A)} Comparison of the maximal capacity as a function of $C$ between indirect translational regulation (ITR, red curves) and direct transcriptional regulation (DTR, black curves) scenarios, for three values of total resource cost (total mRNA number), $\Cost$. Here, $y_{\rm max}=10$, $F=1$. Compared to Fig.~\ref{f2}, ITR stops being beneficial compared to DTR at lower $C$.
{\bf (B)} Capacity difference between the ITR and DTR models as a function of $C$ and $\Cost$. 
The thick red line marks the $(C,\Cost)$ values that lead to equal capacities; in the regime above the red line ITR, in spite of intermediary regulatory steps, is beneficial over DTR.}
\label{f5}
\end{figure*}

\subsection{Comparison at equal resources}
\label{secEqualResources}
In the previous section we compared the optimal operating points for the two regulation scenarios independently, neglecting the resource cost associated with reaching the corresponding optima. However, indirect translational regulation evidently consumes more resources than direct regulation. The extra resource cost is set by $M$, which determines the cost of the mRNA for the intermediary protein $y$, as well as $y_{\rm max}$, which determines the cost of the $y$ protein itself. While it is difficult to convert the cost of mRNA and cost of protein into the same ``currency,'' we can nonetheless compare the costs of the ITR and DTR schemes in terms of the total number of mRNA molecules $\Cost$ alone. 

In the DTR scheme, a good estimate for this cost is $\Cost=N_{\rm max}\gmean$, where $\gmean = \int g(c)P^*(c)dc$ is the expected number of independent output molecules in response to the optimal  distribution of inputs, $P_c^*(c)$. In the ITR scheme, this cost is increased by the number of mRNA molecules needed to implement transcriptional regulation, $M$, such that $\Cost = N_{\rm max}\gmean + M$. In this case, $\gmean$  is implicitly a function of $M$, because different values of $M$ lead to different noise levels and consequently different optimal input distributions $P^*(c)$.

When we fix the total resource cost $\Cost$ in both scenarios, the information in the direct scenario is given by:
\begin{align}
I(c;g)=\log_2 \tilde{Z} + \frac{1}{2}\log_2 N_{\rm max} = \log_2 \tilde{Z} + \frac{1}{2}\log_2 \frac{\Cost}{\gmean} ~. \nonumber\\
\end{align}
For translational repression, the (optimized) information reads
\begin{align}
I(c;g)=\log_2 \tilde{Z}(M^*) + \frac{1}{2}\log_2 \frac{\Cost - M^*}{\gmean} ~, 
\end{align}
where $M^*$ is  the value of $M\in[0,\Cost]$ that optimizes the capacity, reflecting the optimal allocation of available resources, $\Cost$, between the translational mechanism that reduces input noise (favoring large $M$) vs. the reduction of the output noise (favoring large $N_{\rm max}$, i.e. smaller $M$). $M^*$ can be found by first optimizing $\log_2 \tilde{Z}(M)$ for fixed $M$ and then optimizing $I(c;g)$ over $M$ in the second step.

Figure \ref{f5} compares the information capacities of the ITR (red lines) and DTR (black lines) scenarios for three fixed values of the total mRNA number (``cost'') $\Cost$ ($10,100,1000$). Two effects can be observed: First, in both scenarios the capacities increase markedly with $\Cost$; this is expected, because increasing $\Cost$ reduces the overall noise in both scenarios. Note, however, that the gain in capacity upon increasing $\Cost$ is larger in the ITR scenario (compare red curves) than in the DTR scenario (compare black curves), in particular at low $C$. This is also expected because the ITR scheme is more efficient in suppressing input noise; in accordance, optimal mRNA values $M^*$ at low $C$ are higher than at high $C$ (data not shown). Second, due to the additional resource requirements of the ITR scheme, it stops being beneficial over the DTR scheme already at lower $C$, compared to the situation where constrained resources are not taken into account in Fig.~\ref{f2}.

We mapped, systematically, the conditions under which ITR becomes beneficial over DTR. Figure~\ref{f5}B shows the capacity difference $I_{\rm ITR}-I_{\rm DTR}$ as a function of the two main factors that influence capacity, $C$ and $\Cost$. The thick red line marks the combinations $(C,\Cost)$ that lead to equal capacity in both models, i.e. it summarizes the crossing points between ITR and DTR curves in Fig.~\ref{f5}A (thick red points); above the line ITR is superior to DTR, below DTR yields higher capacities. ITR is beneficial at low $C$ (i.e., high input noise), and as $C$ increases, ITR remains beneficial only for resource-intensive regulatory schemes whose cost $\Cost$ grows sufficiently quickly with $C$.   We note, once again, the large size of these information differences, in bits.
%Note that increasing $\Cost$ at fixed low $C$ in the ITR model leads to a larger capacity gain than increasing $C$ at low fixed $\Cost$ by the same order of magnitude in the DTR model.

% Translational regulation scenario consumes more resources than the direct regulation scenario. The extra resource cost is set by both $M$, which determines the cost of the mRNA for the intermediary protein $y$, as well as $y_{\rm max}$, which determines the cost $y$ protein itself. While it is difficult to transform the cost of mRNA and cost of protein into the same ``currency'', we can easily compare the costs of the ITR and DTR schemes in terms of the total number of mRNA molecules alone. Then, the cost of the ITR scheme would be $N_{\rm max} + M$, i.e., the number of independent output mRNA molecules) of $g$ plus the number of mRNA for the intermediary molecules $y$, $M$. In detail, we  fix the total number of mRNA molecules for both scenarios. Information in the direct scenario is then given by $I=\log_2 Z + 0.5\log_2 N_{\rm max}$. For translational repression, the information is $I=\log_2 Z(M^*) + 0.5\log_2 (N_{\rm max}-M^*)$, where $M^*$ is  the value of $M\in[0,N_{\rm max}]$ that optimizes the capacity.

%
%
%
%
%
\section{Discussion}
\label{secDiscussion}
An efficient regulatory pathway will respond to variation in its signal across the full input dynamic range. But what if a part of this dynamic range is associated with very high input noise, as is the case for transcription factor signals at low concentration? Then, the pathway can either avoid responding to those signals completely, thereby sacrificing some of the bandwidth, or it can utilize reaction schemes that are able to reduce the impact of high input noise. In this paper we showed that indirect translational regulation (ITR) is one such regulatory scheme. The intuitive reason for the advantage of translational regulation is that every single one of the $M$ mRNA molecules that is being translationally regulated by the input signal effectively acts as a ``receptor'' for the input concentration; this results in an $M$-fold more efficient averaging of input noise compared to the case of direct regulation, where a single DNA binding site is acting alone as a receptor. Consistent with this intuition, our results show that when input noise is dominant ($C\ll 1$), translational repression with high $M$ can provide large increases in channel capacity relative to direct regulation.

Increases in capacity, however, do not come for free. First, the ITR scheme involves additional reaction steps and  intermediary regulatory molecules; as a consequence, new noise sources, which we call ``ITR noise,'' are introduced. Only when these sources are sufficiently small relative to the input noise set by $C$, does the translational scheme yield measurable benefits. Second, lowering ITR noise sources also incurs metabolic costs associated with producing the required intermediary molecules; this means that a realistic comparison between the direct scheme and the translational repression scheme is relevant only when carried out at comparable resources. We explored these effects in detail to show when translational repression is beneficial over direct regulation.

In this work we have analyzed only one particular reaction scheme for translational regulation, but clearly many variations on the same idea are possible. We examined several of these possibilities, and found consistent results. First, the transformation of the input $c$ into the gene expression level $g$ in the DTR as well as the ITR scheme can be either repressive or activating. We find no qualitative differences between the two schemes, and thus present only the repressive scheme. The channel capacities of the two schemes, separately optimized, differ by $\sim 0.1$ bits, with activation performing slightly better at $C\lesssim 1$, and vice versa. While the shapes of the regulatory curves must be different by construction, the general trends---sharper regulatory curves already for low $C$ in the ITR scheme, lower dissociation constants for high $C$ in the ITR scheme---for the activators are identical to those of the repressors. 

Second, we examined how the precise description of  the input noise form ($\Phi_B$ in Eq.~(\ref{innoise})) influences our results.  In one limit, we can set $\Phi_B = \Phi_1$, corresponding to a single binding site, while in the opposite limits we can set $\Phi_B = \Phi_\infty$, corresponding to a large number of binding sites   \cite{Bialek2005,Kaizu2014,Paijmans2014}.  Tracing through the full numerical analysis in both cases, all differences are small, and largely confined to the regime in which both $C$ and $M$ are small.  The input noise depends on diffusion constants, and we assumed equal diffusion constants for the input and the intermediary molecules, and further  that translationally regulated mRNA are immobile. Relaxing these assumptions changes the effective diffusion rates in the diffusion noise terms; in particular, highly mobile mRNA might further contribute to input noise reduction in the ITR model, Eq.~(\ref{minnoise}). Lastly, we had to make assumptions about how the translationally regulated mRNA are produced and how they interact with the inputs. Apart from a change in ITR noise magnitude, the model where translationally regulated mRNA are present at a molecule count behaves identically to the model we studied, where mRNA are continuously made and degraded.
We have, however, not analyzed more complicated interaction schemes between the input and the mRNA, e.g., one in which the mRNA would sequester the input transcription factors.
%\TRS{Also, we neglect the diffusive binding between mRNA and ribosomes; while this random search process would lead to additional noise contributions analogous to the $y$-input noise in the ITR model, these would arise in the same way both in the ITR and DTR scenarios and therefore not qualitatively change our conclusions.}

What are the limits to noise reduction using translational regulation? In the pedagogical example of the promoter with $M$ regulatory sites, the sites would be packed very closely on the DNA and thus the molecule ``detected'' by one site would have a high likelihood of rebinding to a nearby site, providing a statistically correlated, rather than independent, sample of the concentration in the bulk. In this regime, the decrease in input noise variance would be slower than $1/M$, and as $M\rightarrow\infty$, the input noise would saturate to a bound determined by the linear dimension of the cluster of regulatory sites \cite{Berg1977,Bialek2005,Berezhkovskii2013}. The same argument will ultimately apply to $M$ mRNA molecules of the ITR scenario. There, each mRNA harbors a binding site of size $\ell_c\sim 1-10\e{nm}$, but the mRNAs typically are separated by distances at the nuclear scale $d\sim 1-10\e{\mu m}$. Roughly, we can expect that input readouts will be independent and the ITR mechanism effective so long as $M\lesssim d/\ell_c\sim 10^2-10^4$, putting $M$ in the range that we examined and where it appears biologically plausible. With these $M$, we find significant increases in capacity of $0.8-1.8$ bits at $C\sim 0.1-1$, with optimal regulatory curves using high Hill coefficients $|H^*|\gtrsim 4$, thereby accessing the full output dynamic range. Moreover, the molecular mechanism for integrating the readout over $M$ such ``receptors'' is simple diffusion of the intermediary protein $y$, unlike in the pedagogical example of $M$ regulatory sites on the DNA where we know of no plausible molecular integration mechanism.

Are there biological examples of such indirect translational regulation scheme? During early {\it Drosophila} morphogenesis, one of the primary transcription factors that drive the anterior-posterior (AP) patterning of the embryo is Bicoid, which is established in a graded, exponentially decaying profile along the AP axis \cite{Driever1988a,Driever1988b,Lawrence1992,Gilbert2013,Jaeger2011}.  Bicoid activates a number of downstream genes directly in the anterior and the middle region of the embryo where its concentration is highest.  Absolute concentrations of bicoid are estimated to be $\sim 55\e{nM}$ at a maximum, falling to $\sim 8\e{nM}$ at the midpoint of the embryo \cite{Gregor2007a,Gregor2007b}). At the posterior end of the embryo, where its concentration is low and hard to detect quantitatively using the imaging methods available, Bicoid translationally represses {\it caudal} mRNA \cite{Rivera-Pomar1996,Dubnau1996,Mlodzik1987a,Macdonald1986}. 

The {\it caudal} mRNA molecules are produced by the mother, who deposits them into the egg with a uniform distribution along the AP axis. As they are bound and repressed by Bicoid, while the free ones get translated into Caudal protein, the embryo develops a new gradient of Caudal protein with high concentration at the posterior and low in the anterior. Caudal acts much like the intermediary protein $y$ in our model, becoming a regulator of patterning genes in the posterior, with $c$ representing the Bicoid concentration. While such a scheme involving an intermediary maternal mRNA and protein appears wasteful at first glance, here we have shown that it may provide a substantial benefit for information transmission at low input concentration, i.e. in posterior regions of the embryo. Although we are not familiar with any direct measurement of {\it caudal} mRNA copy numbers, the copy numbers for the gap genes~\cite{Little2013} (and also {\it bicoid}~\cite{Little2011}) are consistent with the range of $M\sim 10^2-10^3$ mRNA per nucleus. More generally, it is interesting to note that {\it bicoid} is just one member of a larger homeodomain TF family whose members have the ability to act both as a transcriptional as well as translational regulators~\cite{Niessing2000}.

\begin{acknowledgments}
We thank T. Gregor, A. Prochaintz, and others for helpful discussions.  This work was supported in part by Grants PHY--1305525 and CCF--0939370 from the US National Science Foundation and by the WM Keck Foundation. AMW acknowledges the support by ERC-Grant MCCIG PCIG10--GA-2011--303561.
\end{acknowledgments}

\appendix
\section{Shot-noise propagation in a generic signaling cascade}
\label{AppendixNoisePropForm}
Let us assume the following signaling cascade, described by a system of Langevin equations for the signaling species $x_j$, 
in which shot noise is generated at each level $j\in [1,..,n]$ and (for $j<n$) propagated into the copy number of species $x_{j+1}$
via a regulatory function $f_{j+1}(x_j)$:
\begin{eqnarray}
\frac{dx_1}{dt}&=&r_1 - \frac{1}{\tau_1}x_1 + \xi_1 \nn \\
&\vdots&\nn\\
\frac{dx_n}{dt}&=&r_n f_n(x_{n-1}) - \frac{1}{\tau_n}x_n  + \xi_n ~, 
\label{eqAppendixCascade}
\end{eqnarray}
In the steady state, for the noise powers of the Langevin noise sources $\xi_j$ we have
\begin{align}
\langle \xi_j(t) \xi_k(t')\rangle = \left( R_j + \frac{\bar{x}_j}{\tau_j} \right) \delta_{jk} \delta(t-t')
				  = \frac{2\bar{x}_j}{\tau_j}\delta_{jk} \delta(t-t') ~,
\label{eqAppendixNoise}
\end{align}
where $R_1 \equiv r_1$ and $R_{j>1} \equiv r_j f_j(x_{j-1})$, and ${\bar x}_j = \langle x_j \rangle$ denotes the stationary mean.
For our purposes we want to compute the overall variance in the last component $\sigma_n^2 = \langle \delta x_n^2 \rangle$.

Linearizing around the means ${\bar x}_j$ via $x_j = {\bar x}_j + \delta x_j$ and $f_j(x_{j-1}) \simeq f_j({\bar x}_{j-1}) + f'_j({\bar x}_{j-1})\delta x_{j-1}$,
and Fourier-transforming for $t \rightarrow \omega$ allows us to convert Eqs.~(\ref{eqAppendixCascade}) into the following equation system for the Fourier-transformed fluctuations $\delta\tilde{x}_j$:
\begin{eqnarray}
-i\omega \delta\tilde{x}_1 &=& -\frac{1}{\tau_1} \delta\tilde{x}_1 + \tilde{\xi}_1 \nn \\
&\vdots&\nn\\
-i\omega \delta\tilde{x}_n &=& r_n f'_n(\bar{x}_{n-1})\delta\tilde{x}_{n-1} - \frac{1}{\tau_n}\delta\tilde{x}_n  + \tilde{\xi}_n ~
\label{eqAppendixCascadeFT}
\end{eqnarray}
For the Fourier-transformed noise powers $\tilde{\xi}_j$ we get
\begin{align}
\langle \tilde\xi_j(\omega) \tilde\xi_k^*(\omega)\rangle = 2\tau_j^{-1}\bar{x}_j\delta_{jk}.
\label{eqAppendixNoiseFT}
\end{align}

The system defined by Eqs.~(\ref{eqAppendixCascadeFT}) can be solved algebraically by succesively inserting the solution for $\delta\tilde{x}_j$ into the equation for $\delta\tilde{x}_{j+1}$:
\begin{eqnarray}
 \delta\tilde{x}_1 &=& \frac{\tilde{\xi}_1}{\tau_1^{-1} - i\omega}	\nn\\
		   &\vdots&						\nn\\
 \delta\tilde{x}_n &=& \sum_{j=1}^n \frac{\tilde\xi_j}{\tau_j^{-1} - i\omega} \prod_{q=j+1}^n \frac{r_q f'_q}{\tau_q^{-1} - i\omega}
\end{eqnarray}
where we abbreviate $f'_q \equiv f'_q(\bar{x}_{q-1})$.

We can now obtain the variance $\sigma_n^2$ by integrating over the noise power spectrum $S_n(\omega)=\langle \delta\tilde{x}_n \delta\tilde{x}_n^* \rangle$ of the fluctuations in component $n$:
\begin{align}
 \sigma_n^2	&= \int \frac{d\omega}{2\pi} S_n(\omega) = \int \frac{d\omega}{2\pi} \langle \delta\tilde{x}_n \delta\tilde{x}_n^* \rangle \nn\\
		&= \int \frac{d\omega}{2\pi} \sum_{j=1}^n \frac{2\bar{x}_j \tau_j^{-1}}{\tau_j^{-2} + \omega^2} \prod_{q=j+1}^n \frac{\left(r_q f'_q\right)^2}{\tau_q^{-2} + \omega^2}
\end{align}
where we recall that for mixed indices $\langle \tilde\xi_j(\omega) \tilde\xi_k^*(\omega)\rangle = 0$.

Isolating the $n$-term of the sum, and introducing the dimensionless integration variable $w \equiv \tau_n\omega$, we can further write:
\begin{align}
 \sigma_n^2	&=\bar{x}_n \underset{=1}{\underbrace{\int\frac{dw}{\pi} \frac{1}{1+w^2}}}		\nn\\
		& \quad\quad + \sum_{j=1}^{n-1} \int \frac{dw}{\pi} \frac{\bar{x}_j \left(\frac{\tau_j}{\tau_n}\right)}{1 + \left[\frac{\tau_j}{\tau_n}w\right]^2} 
					      \prod_{q=j+1}^{n} \frac{\left(r_q \tau_q f'_q\right)^2}{1 + \left[\frac{\tau_q}{\tau_n}w\right]^2}	\nn\\
\end{align}
%The first integral can be directly evaluated as shown.
The integrand of the second integral can be written as:
\begin{align}
%\prod_{k=j}^{n} \left(1+\left[\frac{\tau_k}{\tau_n}w\right]^2\right)^{-1} = \left(1+w^2\right)^{-1} \prod_{k=j}^{n-1} \left(1+\left[\frac{\tau_k}{\tau_n}w\right]^2\right)^{-1}
\prod_{k=j}^{n} \left(1+\left[\frac{\tau_k}{\tau_n}w\right]^2\right)^{-1} = \frac{1}{1+w^2} \prod_{k=j}^{n-1} \left(1+\left[\frac{\tau_k}{\tau_n}w\right]^2\right)^{-1}
\end{align}
The leading factor $(1+w^2)^{-1}$ only has significant contributions when $|w| \lesssim 1$,
and will be suppressed (together with the whole integrand) for $|w|\gg 1$.
Assuming that $\tau_n$ is the longest timescale of the problem, i.e. $\forall k: \tau_n \gg \tau_k$, 
in the relevant regime $|w|\lesssim 1$ all factors except for the leading one will be $\simeq 1$.
Thus, to a good approximation:
\begin{align}
 \sigma_n^2	&\simeq \bar{x}_n + \sum_{j=1}^{n-1} \bar{x}_j \left(\frac{\tau_j}{\tau_n}\right) \prod_{q=j+1}^n \left(r_q \tau_q f'_q\right)^2 \int \frac{dw}{\pi} \frac{1}{1+w^2}	\nn\\
		&= \bar{x}_n + \sum_{j=1}^{n-1} \underset{\equiv\sigma^2_{n\leftarrow j}}{\underbrace{ \left[ \prod_{q=j+1}^n r_q \tau_q f'_q \right]^2 \times \left(\frac{\tau_j}{\tau_n}\right) \times \bar{x}_j }}
\end{align}

%
%
%
%
%
%%% BIBLIOGRAPHY %%%
% Make sure folder 'Bib' is present to build this correctly
\bibliography{Bib/General,Bib/InfoTransmission,Bib/GapGenes,Bib/BicoidPreSteadyStateDecoding}
\end{document}